\documentclass[10pt]{article}
\usepackage{amsmath,amssymb,amsfonts,epsfig,graphics,graphicx}
\newcommand{\be}{\begin{equation}}
\newcommand{\ee}{\end{equation}}
\newcommand{\ea}{\end{eqnarray}}
\newcommand{\ba}{\begin{eqnarray}}

\newcommand{\Avec}{{\bf A}}

\begin{document}

\title{\Large Vortex Solutions and a Novel Role for R-parity in an $N$=2-Supersymmetric Model for Graphene}

\author{Everton M. C. Abreu$^{1}$ \thanks{E-mail: evertonabreu@ufrrj.br}$\;$, Marco A. De Andrade
$^{2,5}$ \thanks{E-mail: marco@cbpf.br}$\;$, Leonardo P. G. De Assis$^{6}$ \thanks{E-mail: lpgassis@stanford.edu}$\;$, \\
Jos\'{e} A. Helay\"{e}l-Neto$^{3,5}$ \thanks{E-mail: helayel@cbpf.br}$\;$,  
A. L. M. A. Nogueira$^{4,5}$ \thanks{E-mail: nogue@cbpf.br}$\;$, and Ricardo C. Paschoal$^{4,5}$ \thanks{E-mail: paschoal@cbpf.br}  \\  \\
$^{1}${\it  \normalsize Departamento de F\'{\i}sica, Universidade Federal Rural do Rio de Janeiro,} \\
{\it \normalsize BR 465-07, 23890-971, Serop\'edica, RJ, Brasil.}\\
$^{2}${\it \normalsize Universidade do Estado do Rio de Janeiro (Resende-RJ),} \\
{\it \normalsize Rodovia Presidente Dutra, km 298, P\'olo Industrial, CEP 27537-000, Resende, RJ, Brasil.}\\
$^{3}${\it  \normalsize Centro Brasileiro de Pesquisas F\'{\i}sicas -- CBPF,} \\
{\it \normalsize Rua Dr. Xavier Sigaud 150, 22290-180, Rio de Janeiro, RJ, Brasil.}  \\
$^{4}${\it  \normalsize Centro Federal de Educa\c{c}\~ao Tecnol\'ogica Celso Suckow
da Fonseca (CEFET/RJ), } \\
{\it \normalsize Av. Maracan\~{a}, 229, 20271-110, Rio de Janeiro, RJ, Brasil.} \\   
$^{5}${\it \normalsize Grupo de F\'{\i}sica Te\'{o}rica Jos\'e Leite Lopes,} \\
{\it \normalsize P.O.\ Box 91933, 25685-970, Petr\'opolis, RJ, Brasil.} \\   
$^{6}${\it \normalsize Stanford University,} \\
{\it \normalsize CSLI, Ventura Hall, 220 Panama Street, Stanford, CA 94305-4101.}  }

\date{April 23 2014}
\maketitle

\begin{abstract}
\noindent In a previous work, we have been able to settle Jackiw's et al.\ chiral gauge theory for Dirac fermions in graphene in an
$N$=1 supersymmetric framework, using a $\tau_{3}$-QED prescription, defined by means of a single pair of gauge charged superfields, but without preserving a global phase symmetry associated to the electric charge. In the present work, we propose another $N$=1-generalisation which indeed preserves this symmetry, namely, a straightforward extension built upon a set of two pairs of (chiral) gauge-charged superfields plus an extra pair of electrically neutral superfields. 
We then further proceed to establish, via a dimensional reduction procedure, an $N$=2 extension, allowing for the identification of non-perturbative features, as we put forward Bogomol'nyi equations and obtain vortex-like solutions saturating a topologically non-trivial bound. Remarkably, the bosonic projection of the $N$=2 functional space onto the saturated regime analysed herewith reveals to be free from extra scalar degrees of freedom that would otherwise demand a phenomenological interpretation. 
The investigation of Jackiw's model within an $N$=2 complex superspace is also motivated
by the assumption that an R-parity-like symmetry could provide a route to incorporate the global phase-fermion
number invariance as an external-like symmetry of the theory, thus associating the electric charge in graphene to the complex covariance (super-)space for the $N$=2-$D$=3 setup.
We prove such a hypothesis to be realisable, as we build up the model endowed with all the symmetries required to further extend Jackiw's chiral gauge theory.  
\end{abstract}

\newpage

\section{Introduction}
Although theoretically studied since 1947~\cite{Wallace}, graphene~\cite{Katsbook} was produced for the first time only in 2004~\cite{Nov1, Nov2, Nov3, Kim}. It is a monolayer, two-dimensional honeycomb array of carbon atoms and shows very interesting properties~\cite{Int1, Int2, Int3, Int4, Int5, Rev1, Rev2, Rev3, Rev4, Rev5, Rev6, Rev7, Rev8, Rev9, Rev10, CN2010a, CN2010b, Peres2010, DasSar2011, Goer2011, Kotov, GeimNobel, KostyaNobel, Rev11}, many of which are condensed-matter manifestations of identical ones in high energy physics (HEP)~\cite{Sem, Frad, Hal, RevRel1, Been2008, RevRel2, VKG}. Furthermore, similarities with other equally interesting, new materials such as topological insulators~\cite{HK, QZ, BHZ} have been found~\cite{CGV, MBM}, with potential applications in spintronics or quantum computation~\cite{giant}.
In this line of development, the emergence of Majorana fermions~\cite{Nat-Wilc, MF-trig, MF1, MF2, MF3, Mourik, Leij, MF4, Stanescu, MF-Graph-like} in graphene is a theoretical perspective, although not yet experimentally realised. 

Among the many aspects of HEP related to graphene, we extensively use here the gauge and scalar fields that emerge from deformations or inhomogeneities of the honeycomb lattice~\cite{Rev1, Rev2, Rev3, Rev4, Rev5, Rev6, Rev7, Rev8, Rev9, Rev10, RevRel1, gauge1, gauge2, gauge3, strain}\footnote{One can also consider additional \emph{external} gauge fields, if the model is meant to describe, for instance, a magnetic field applied to a graphene sample, a setup that may generate interesting features associated to zero mode vortex configurations~\cite{BitanRoy}.}, in addition to the Dirac fermion that describes the low energetic electrons in graphene. As we have already done in Ref.~\cite{nosso}, the chiral model proposed by Jackiw et al.~\cite{jp,jp2,jp3} is given special attention. In addition, we propose supersymmetry (SUSY), as we have already done in Ref.~\cite{nosso}, but now with emphasis on analysing the $N$=2-extension and on the realisation of the full set of symmetries defined in Jackiw's et al.\ theoretical setup. With this purpose, we discuss the invariances of our $N$=2-supersymmetric model, paying special attention to the (local) Jackiw's chiral gauge symmetry and the global phase symmetry, the latter initially proposed by Jackiw and co-authors to be associated to the electric charge of the electron. In this paper, we show that the global phase symmetry can be realised through an inherent $N$=2-superspace-structure R-parity-like invariance, thus establishing an interesting connection between the conservation of the electric charge and an external (super-)space symmetry. 

We also perform a phenomenological approach, as we work on finding vortex solutions. 
The relationship between SUSY and BPS-vortex solutions was proven in Refs.~\cite{OliveWitten, HS1, HS2, HS3}, where Bogomol'nyi equations emerge if half of the supersymmetries have the corresponding generator action onto the fields leading to a null result. By these means, we succeed to find vortices in graphene, as shown in detail below. These solutions are a good indication for our model, since there are many works in graphene literature that also predict vortices, the most proeminent one, in this case, being exactly the abovementioned Jackiw's et al.\ model. Thus, as shown in the present work, SUSY may also be viewed as a possible theoretical mechanism on its own merit, to be used with the simple purpose of obtaining vortex solutions without any need of an a priori interpretation of superpartner fields. Interestingly enough, for the particular BPS state obtained in this work, extra bosonic superpartners inherent to the SUSY-extension are shown to vanish. Experimental tests of vortex solutions are attainable by means of available techniques, sometimes referred to as `artificial' graphene, which allow to implement Dirac fermions as well as the gauge and scalar fields by means of other base material. Those experimental setups are: ultracold atom optical-lattice graphene~\cite{optlatt} and STM-assembled molecular graphene~\cite{molecgraph}. In many cases, it is easier to control and deal with these materials than with graphene itself. For example, in Ref.~\cite{vortCu}, a vortex solution is proposed in detail for the molecular graphene constructed in Ref.~\cite{molecgraph}.

The present work is organised as follows: we shall review, in the next Section, some basic facts about the
mentioned chiral gauge theory and, in Section~3, we shall set up the main aspects of its $N$=1 supersymmetric
generalisation, as done in the work of Ref.~\cite{nosso}. We then present an alternative $N$=1 SUSY model augmented
by an extra pair of $N$=1 scalar superfields, and we discuss the features that distinguish the two proposals. Then, in Section~4, we shall explicitly build up an $N$=2-supersymmetric action in (2+1)-dimension space-time, obtained as the result
of a dimensional reduction procedure performed upon a suitable $N$=1-D=3+1 supersymmetric theory. We also present the proper R-Parity prescription that plays the role of the fermion number/electric charge global symmetry. We conclude Section~4 by exhibiting the $N$=2-D=3 SUSY functional that extends the action of Ref.~\cite{jp}, with full realisation of the complete set of proposed symmetries. In Section~5, Bogomol'nyi equations and corresponding vortex solutions are obtained, and numerical simulations are presented. Finally, conclusions and perspectives are depicted in the last Section.
%
\vspace{1cm}

\section{Jackiw's chiral gauge theory for graphene}

In 2000, Chamon studied in detail~\cite{cham2000} some consequences of the so-called Kekul\'e distortion in a honeycomb array of carbon atoms. In 2007, Hou, Chamon and Mudry (HCM) extended this idea~\cite{hcm} considering a Kekul\'e \emph{texture}, that is, a different Kekul\'e distortion in each point of the plane, thus introducing a scalar field, which will be represented here by $\varphi$. Some time later, in order to provide dynamics and finite energy to the vortices described by HCM, Jackiw and Pi~\cite{jp} introduced a gauge field $A_\mu$ to their model, thus formulating the chiral gauge theory that will be summarised in the present Section. For more details about this theory and its extensions~\cite{jp2,jp3}, the reader is referred to the original articles.

Jackiw's et al.\ model is described by the following Lagrangian density:
\begin{equation}
\mathcal{L}_{\mathrm{HCM-JP}}=\overline\psi_{+}\gamma^{\mu}(iD_{\mu}^{+})\psi_{+}
+ \overline\psi_{-}\gamma^{\mu}(iD_{\mu}^{-})\psi_{-} - g\varphi\overline
\psi_{+}\psi_{-} - g\varphi^{\ast}\overline\psi_{-}\psi_{+} ~,\label{LHCM2}%
\end{equation}
with $\mu$ spanning the values (0,1,2); the gamma matrices given by $\gamma^{\mu}=(\sigma_{z},
i\sigma_{y}, -i\sigma_{x})$, with $\sigma_j$ ($j=x, y, z$) being the usual Pauli matrices; $\psi_{+}$ and $\psi_{-}$ being the two-component spinors describing electrons in graphene; $\varphi$, the complex scalar field describing the Kekul\'e texture; and the covariant derivatives $iD_{\mu}^{+}=i\partial_{\mu}-qA_{\mu}~$ and $~iD_{\mu}^{-} =i\partial_{\mu}+qA_{\mu}$, where $A_\mu$ is the gauge field and $g$ and $q$ are coupling constants. The model is invariant under the following local chiral transformation:
\begin{equation}
\varphi\to e^{2 i q \omega}\, \varphi\;, \qquad\Psi\to e^{i q \omega\gamma_{5}%
^{J}}\, \Psi\;, \qquad\Avec \rightarrow\Avec +
\nabla\,\omega\;;\label{rjeq9}
\end{equation}
\begin{equation}
\Psi_{\pm}\to e^{\pm iq \omega}\ \Psi_{\pm}, \qquad\bar{\Psi}_{\pm}\to\bar
{\Psi}_{\pm}\, e^{\mp iq \omega} ,\label{rjeq12}
\end{equation}
where $\psi_{-}\equiv\gamma^{0}\psi^{\prime}_{-}$, 
\begin{equation}%
\begin{array}
[c]{lcr}%
\Psi= \left(
\begin{array}
[c]{c}%
\psi_{+}\\[1ex]%
\psi^{\prime}_{-}%
\end{array}
\right) ,~ & \Psi_{+} = \left(
\begin{array}
[c]{c}%
\psi_{+}\\[1ex]%
0
\end{array}
\right) ,~ & \Psi_{-} = \left(
\begin{array}
[c]{c}%
0\\[1ex]%
\psi^{\prime}_{-}%
\end{array}
\right) 
\end{array},
\label{Psis}%
\end{equation}
and
\be
\gamma_5^{J} =  {\binom{I\qquad0 }{0\quad
-I}} , 
\ee
with $I$ being the $2\times 2$ identity matrix.

Also, a discrete Parity symmetry prevails:
\begin{eqnarray}
\Psi \; \xrightarrow{\mbox{\tiny{Parity}}} \;\; i \gamma^{3}\gamma^{1}\; 
\Psi & = & \left( \begin{array}{cc} \sigma_{y} & 0 \\ 0 & \sigma_{y} \end{array} \right) \; \Psi \; .
\end{eqnarray}

Moreover, this system possesses a global fermion number symmetry, with just
the Fermi fields transforming with a constant phase: $\Psi\to e^{i\alpha}\,
\Psi$. Consequently, as for the set of continuous symmetries, one can list a local chiral $U(1)$ and a
global $U(1)$ fermion number invariances. At this point, we would like to stress that the gauged Abelian symmetry does {\it not} represent the usual interaction between electrically charged matter and the gauge boson. As a matter of fact, it is the additional global Abelian phase symmetry that happens to be related to the electric charge. Because the theory resides in $(2+1)$
dimensions, no chiral anomalies interfere with the chiral gauge symmetry.

%

\section{\bigskip $N$=1-D=1+2-Supersymmetric Actions}

Adopting, for the $N$=1-D=1+2 superspace, the usual ($x^{\mu},\theta$) parameterisation,  where $x^{\mu}$ are the coordinates of the
D=1+2 spacetime and the fermionic coordinates, $\theta$, are Majorana spinors,
$\theta^{c} =\theta$, we define the complex scalar $N$=1, D=(2+1)-superfields with opposed
$U(1)$-charges, $\Phi_{+}$ and $\Phi_{-}$, as
\begin{equation}
\Phi_{\pm}=A_{\pm}+\overline{\theta}\psi_{\pm}-\frac{1}{2}\overline{\theta
}\theta F_{\pm}~\text{ and }~\Phi_{\pm}^{\dagger}=A_{\pm}^{\ast}%
+\overline{\psi}_{\pm}\theta-\frac{1}{2}\overline{\theta}\theta F_{\pm}^{\ast}
~,\label{scalar3}%
\end{equation}
where $A_{\pm}$ are complex scalars, $\psi_{\pm}$ are Dirac spinors and
$F_{\pm}$ are auxiliary scalar fields that obey the following supersymmetry
transformations:
\begin{align}
\delta A_{\pm}  & = \overline{\varepsilon}\psi_{\pm}\\
\delta\psi_{\pm}  & = \varepsilon F_{\pm}+i\varepsilon\gamma^{\mu}%
\partial_{\mu}A_{\pm}\\
\delta F_{\pm}  & = i\,\overline{\varepsilon}\gamma^{\mu}\partial_{\mu}%
\psi_{\pm}~.
\end{align}

Notice that, by introducing $A_\pm$ and $\psi_\pm$ as components of the same superfield, the scalars cannot be neutral under the global phase - $U(1)$ transformation~\cite{jp,jp3} if such a symmetry is imposed at the (susy-covariant) level of superfields. Thus, one faces two alternative strategies to implement the electric charge related phase invariance: either one redefines the phase transformation by means of the superspace structure, if such a redefinition is indeed achievable, or one has to supplement the functional space with (at least) an extra superfield. This additional complex scalar susy-multiplet must then be invariant under the global phase transformation, leaving its zero-$\theta$ complex scalar component-field invariant as well. We now present a summary of the $N$=1 framework upon which the supersymmetric extension of Jackiw-Pi's chiral gauge theory was worked out in Ref.~\cite{nosso}, and, after exhibiting that model, we turn to another $N$=1 generalisation of Jackiw-Pi's construction, obtained in a straightforward approach, by means of introducing {\it a pair} of additional superfields. 

In the Wess-Zumino gauge, the gauge superconnection, $\Gamma_{a}$, is written
as
\begin{equation}
\Gamma_{a}=i\left(  \gamma^{\mu}\theta\right) _{a}A_{\mu}+\overline{\theta
}\theta\lambda_{a}~\text{ and }~\overline{\Gamma}_{a}=-i\left(  \overline
{\theta}\gamma^{\mu}\right) _{a}A_{\mu}+\overline{\theta} \theta
\overline{\lambda}_{a} ~,\label{gauge3}%
\end{equation}
where $A_{\mu}$ is the gauge boson and $\lambda_{a}$ is its partner, the gaugino (Majorana
spinor), with the spinorial index $a$ ranging from 1 to 2. Defining the ``field-strength'' superfield $W_{a}$ as
\begin{equation}
W_{a}=-\frac{1}{2}\overline{D}_{b}D_{a}\Gamma_{b}~,
\end{equation}
with covariant derivatives given by
\begin{equation}
D_{a}=\overline{\partial}_{a}-i\left(  \gamma^{\mu}\theta\right)  _{a}%
\partial_{\mu}~\text{ and }~\overline{D}_{a}=-\partial_{a}-i\left(
\overline{\theta}\gamma^{\mu}\right) _{a}\partial_{\mu}~,
\end{equation}
where $\partial_{a} = \partial/\partial\theta^a$ and $\overline{\partial}_{a} = \partial/\partial\overline{\theta}^a$, one obtains
\begin{equation}
W_{a}=\lambda_{a}+\Sigma^{\mu\nu}_{ab}\theta_{b}F_{\mu\nu}-\frac{i}
{2}\overline{\theta}\theta\gamma^{\mu}_{ab}\left( \partial_{\mu} \lambda
_{b}\right) \label{strenght3a}%
\end{equation}
and
\begin{equation}
\overline{W}_{a}=\overline{\lambda}_{a}-\overline{\theta}_{b}\Sigma^{\mu\nu
}_{ab}F_{\mu\nu}-\frac{i}{2}\overline{\theta}\theta\left( \partial_{\mu}
\overline{\lambda}_{b}\right) \gamma^{\mu}_{ab}~,\label{strenght3b}%
\end{equation}
where $\Sigma^{\mu\nu}=\frac{1}{4}\left[  \gamma^{\mu},\gamma^{\nu}\right]  $
are the generators of the Lorentz group in (2+1) space-time dimensions and, as usual, $F_{\mu\nu}=\partial_\mu A_\nu - \partial_\nu A_\mu$ is the gauge-field strength.

The gauge covariant derivatives that act on the matter fields with opposed
$U(1)$-charges, $\Phi_{+}$ and $\Phi_{-}$, are respectively given by
\begin{equation}
\nabla\Phi_{\pm}=\left(  D_{a}\mp iq\Gamma_{a}\right)  \Phi_{\pm}~\text{ and
}~\overline{\nabla}\Phi_{\pm}^{\dagger}=\left(  \overline{D}_{a}\pm
iq\overline{\Gamma}_{a}\right)  \Phi_{\pm}^{\dagger} ~.\label{deriv3}%
\end{equation}

Using the definitions previously given for the superfields in
Eqs.~(\ref{scalar3}), (\ref{gauge3}), (\ref{strenght3a}), (\ref{strenght3b})
and the gauge covariant derivatives given in Eq.~(\ref{deriv3}), one can build the supersymmetric $\tau_{3}$-QED action~\cite{DelCima-Marco1, DelCima-Marco2},
\begin{equation}
S_{\tau_{3}\text{-QED}}= {\displaystyle\int} d^{3}xd^{2}\theta\left\{
-\frac{1}{2}\overline{W}W+(\overline{\nabla}\Phi_{+}^{\dagger})(\nabla\Phi
_{+})+(\overline{\nabla}\Phi_{-}^{\dagger})\left( \nabla\Phi_{-}\right)
+2m\left( \Phi_{+}^{\dagger}\Phi_{+}-\Phi_{-}^{\dagger}\Phi_{-}\right)
\right\}  ~,\label{superqedtau3}%
\end{equation}
and suplemment it with 
the most general local $U(1)$ and parity-invariant
\footnote{Parity transformation action on superfields: $\Phi_{\pm} \rightarrow - \Phi_{\mp}$.} supersymmetric $\Phi^{4}$-sector action:
\begin{eqnarray}
S_{\Phi^{4}}&=& f {\displaystyle\int} d^{3}xd^{2}\theta\left[\left( \Phi_{+}^{\dagger}\Phi_{+}\right)^2 - \left(\Phi_{-}^{\dagger}\Phi_{-}\right)^2 \right] + \nonumber\\
 & & + h{\displaystyle\int} d^{3}xd^{2}\theta\left[ 
\left(\Phi_+^{\dag}\Phi_-\Phi_+\Phi_+ + \Phi_+\Phi_-^{\dag}\Phi_+^{\dag}\Phi_+^{\dag}\right) 
-\left( \Phi_-^{\dag}\Phi_+\Phi_-\Phi_- + \Phi_-\Phi_+^{\dag}\Phi_-^{\dag}\Phi_-^{\dag} \right)
  \right] \label{superphi4} \\
  & \equiv & S_f + S_h, \nonumber
\end{eqnarray}
where $f$ and $h$ are (real) coupling constants (to be associated later with Jackiw-Pi's $g$ present in Eq.~(\ref{LHCM2})).
The quartic term with coupling constant $h$ explicitly breaks the global $U(1)$-phase fermion number symmetry. Nevertheless, this term may be kept in order to investigate possible regimes where the chiral interaction dominates over the forces dictated by the global symmetry. The chiral symmetry has a dynamical character in that it dictates a gauge interaction; on the other hand, the global fermion number appears to have only a kinematic character. The latter classifies the states and field configurations (e.g., vortices in Refs.~\cite{jp,jp3}) without, however, introducing gauge-type interactions. Nothing prevents us from setting $h=0$ whenever we wish to recover results for which fermion number conservation is mandatory, but we wish to avoid such a restriction to the superpotential.

It was shown in Ref.~\cite{nosso} that the model summarised above produces the following component-wise expression for the spinor-scalar 3- and 4-vertex sectors of the complete supersymmetric action:
\begin{eqnarray}
&& S_{\mbox{\tiny sp-sc int.}}\,=\, \nonumber \\
&=& {\displaystyle\int} d^{3}x \left\{ - i q \biggr(A_{+} \overline\psi_{+}\lambda- A_{-}\overline\psi_{-}\lambda-
A_{+}^{\ast}\overline\lambda\psi_{+} + A_{-}^{\ast}\overline\lambda\psi_{-}
\biggr) \, -  \left[ 2f {|A_{+}|}^{2} + h \left( A_{+}A_{-} + A_{+}^{*}A_{-}^{*} \right)\right] \, {\overline{\psi}}_{+}\psi_{+} \right.  \nonumber \\
&+& \left[ 2f {|A_{-}|}^{2} + h \left( A_{+}A_{-} + A_{+}^{*}A_{-}^{*} \right)  \right]{\overline{\psi}}_{-}\psi_{-} \, - \, \left[ \frac{f}{2} {A_{+}^{*}}^{2} + \frac{h}{2} A_{+}^{*}A_{-} \right] \overline{{\psi}_{+}^{c}}\psi_{+} \, - \, \left[    \frac{f}{2} A_{+}^{2} + \frac{h}{2} A_{+}A_{-}^{*}\right]{\overline{\psi}}_{+}\psi_{+}^{c}  \nonumber \\
&+& \, \left[ \frac{f}{2} {A_{-}^{*}}^{2} + \frac{h}{2} A_{-}^{*}A_{+} \right] \overline{{\psi}_{-}^{c}}\psi_{-} \, + \, \left[    \frac{f}{2} A_{-}^{2} + \frac{h}{2} A_{-}A_{+}^{*}\right]{\overline{\psi}}_{-}\psi_{-}^{c} \, - \, h\,|A_{+}|^{2}\left( \overline{{\psi}_{+}^{c}}\psi_{-} + {\overline{\psi}}_{-}\psi_{+}^{c} \right)  \nonumber \\ 
&+& \left. \, h\,|A_{-}|^{2}\left( \overline{{\psi}_{-}^{c}}\psi_{+} + {\overline{\psi}}_{+}\psi_{-}^{c} \right) \, - \frac{h}{2}\left( A_{+}^{2}  -  {A_{-}^{*}}^{2} \right) {\overline{\psi}}_{+}\psi_{-} \, - \frac{h}{2}\left( {A_{+}^{*}}^{2}  -  A_{-}^{2} \right) {\overline{\psi}}_{-}\psi_{+} \right\}. \nonumber 
\end{eqnarray}

The content of the above Lagrangian density together with the fermionic minimal couplings with the gauge boson 
that emerges in the component-wise version of Eq.~(\ref{superqedtau3}) demonstrate that the supersymmetric $\tau_{3}$-QED theory given
in Eq.~(\ref{superqedtau3}), supplemented by a $\Phi^{4}$-term given in
Eq.~(\ref{superphi4}), provides a theoretical framework that extends Jackiw-Pi's original chiral gauge theory~\cite{jp}, represented here by Eq.~(\ref{LHCM2}), {\it whenever one gives up the global phase symmetry}. After some identifications, for example, in a crude comparison, one would associate Jackiw-Pi's scalar field $\varphi$ with local-$U(1)$ doubly charged combinations of squares of the type $A_{+}^{2}$, $A_{-}^{2}$ and complex conjugates, which naturally arise from our quartic superfield action, Eq.~(\ref{superphi4}). Nevertheless, such combinations face the issue of having undefined global phase transformation rule, whenever one attempts to implement the symmetry exclusively at the level of superfields, as we have anticipated. In other words, a global phase variation of $\Phi_{\pm}$ would imply that the scalar field $A_{\pm}$ should transform evenly with the corresponding fermions $\psi_{\pm}$, which in their turn should vary with the same phase (as both $\psi_{+}$ and $\psi_{-}$ compose the description of the electron). As it goes, a combination of the kind ${A_{+}^{*}}^{2}  -  A_{-}^{2}$ renders the scalar field in the Yukawa term undefined w.r.t. the global phase transformation. Furthermore, to exactly identify terms and degrees of freedom one should establish the proper combinations of fermions that diagonalise the mass matrix (eigenvectors) upon a particular choice of scalar fields configuration that minimises the potential\footnote{Indeed, this was done in Ref.~\cite{nosso}, but unfortunately a mistake was made by us with respect to the values of the fermion masses in the published version (they are right in the preprint informed in Ref.~\cite{nosso}, and no consequences to the conclusions are implied), hence we would like to stress here the correct information: the five fermion masses values after diagonalisation are not the three degenerate null values and the two degenerate others with value $-m$ as stated in the published version, but, instead, they are those informed in the preprint version, namely: two denerate null outcomes; one $-m$ value; and two masses given by $-\frac{m}{2}\left( 1 \pm \sqrt{2-\frac{4q^2}{mf}}\right)$, where the assumption $h=0$ was explicitly made in this particular calculation.}. 

\subsection{An Alternative Approach}
As an alternative to the $N$=1-supersymmetric $\tau_{3}$-QED approach, we now present a straightforward $N$=1-generalisation of Jackiw-Pi's chiral gauge theory~\cite{jp}. The guiding idea is to implement the original symmetries in the minimum possible set of operators. One is then commited to the susy extension of a minimal coupling matter-gauge interaction plus a Yukawa term, where the scalar degrees of freedom coupled to fermions should regard the global phase transformation as an identity operation. With this purpose, one would introduce a single extra complex scalar superfield meant to host a global phase invariant scalar. However, the requirement of invariance under parity symmetry and the reality constraint impose a {\it pair} of complex scalar superfields as a solution to a full symmetric action\footnote{A single extra superfield $\Omega$ would lead the quartic interaction that is meant to reproduce Jackiw-Pi's Yukawa term to the inconsistent mapping, under Parity, of an operator of the kind $\Phi_{+}^{\dagger}\Phi_{-}\Omega^{2}$ onto its hermitian conjugate and vice-versa. As the superpotential must be odd with respect to Parity - the integration measure being odd as well, the superpotential would be an imaginary quantity, and an $i$ correcting factor would jeopardise the generalisation of Jackiw-Pi's model. On the other hand, one could consider a 3-vertex proposal, realised through the superpotential $h^{\prime}\left(\Phi_{+}^{\dagger}\Phi_{-}\Omega \, + \, \Phi_{-}^{\dagger}\Phi_{+}\Omega^{\dagger}\right)$, where a single extra $\Omega$-superfield is Parity-transformed onto $-{\Omega}^{\dagger}$. Nevertheless, as we consider a $(2+1)$-dimensional model, a 3-vertex superpotential would allow to regard the coupling constant $h^{\prime}$ as a v.e.v of a field, with yet undefined transformation properties, a possibility that we have chosen to avoid so far. }:
\begin{equation}
\Omega_{\pm} =\phi_{\pm} +\overline{\theta}\omega_{\pm} -\frac{1}{2}\overline{\theta
}\theta S_{\pm}~\text{ and }~\Omega_{\pm}^{\dagger}=\phi_{\pm}^{\ast}%
+\overline{\omega}_{\pm}\theta-\frac{1}{2}\overline{\theta}\theta S_{\pm}^{\ast}
~,\label{omega}
\end{equation}
where the ${\pm}$-indices carried by the $\Omega$ superfields refer to the same $U(1)$-gauge symmetry charges associated to the $\Phi_{\pm}$ supermultiplets. Concerning the components, $\phi_{\pm}$ are complex scalar fields, $\omega_{\pm}$ are Dirac spinors and
$S_{\pm}$ are auxiliary scalar fields, at the very same footing of the component fields in $\Phi_{\pm}$. Parity also acts on $\Omega_{\pm}$  in strict analogy with its effect on $\Phi_{\pm}$:
\begin{equation}
\Omega_\pm \,\rightarrow \, -\,\Omega_\mp \label{spomegaparity}\, ,
\end{equation}

As for the global phase symmetry, the $\Phi$-superfields transform according to $\Phi_{\pm}^{\prime}\, =\, e^{i\alpha}\Phi_{\pm}$, so sharing the same ``electric charge"\footnote{One should remark that as far as the fields $\psi_{+}$ and $\psi_{-}$ are taken as solutions of a {\it formally} relativistic Dirac equation, both particle (electron) and anti-particle (``hole") degrees of freedom are considered in this work. On this token, from a field-theoretical point of view, we keep the assignment ``electron field", with all implications that arise from a Dirac equation. A complete approach to proper identification of electron and ``hole" degrees of freedom, described upon the basis spanned by $\psi_{+}$ and $\psi_{-}$, as well as comments on the connection between charge and (pseudo)spin in (2+1)D can be found, e.g., in the work of Ref.~\cite{Katsbook}.}. The $\Omega$-superfields, as already discussed, remain invariant.

The $N$=1-D=1+2 susy action follows:
\begin{eqnarray}
S_{\text{min.}} &=& {\displaystyle\int} d^{3}xd^{2}\theta\left\{
-\frac{1}{2}\overline{W}W+(\overline{\nabla}\Phi_{+}^{\dagger})(\nabla\Phi
_{+})+(\overline{\nabla}\Phi_{-}^{\dagger})\left( \nabla\Phi_{-}\right) +(\overline{\nabla}\Omega_{+}^{\dagger})(\nabla\Omega_{+})  +(\overline{\nabla}\Omega_{-}^{\dagger})\left( \nabla\Omega_{-}\right) + \nonumber \right. \\ 
&&\left. +h\left[\left( \Phi_{+}^{\dagger}\Phi_{-}\Omega_{+}^{2}+\Phi_{-}^{\dagger}\Phi_{+}{\Omega_{+}^{\dagger}}^{2}\right) - \left( \Phi_{+}^{\dagger}\Phi_{-}{\Omega_{-}^{\dagger}}^{2}+\Phi_{-}^{\dagger}\Phi_{+}\Omega_{-}^{2}\right)
\right]\right\}  ~.\label{minimum}%
\end{eqnarray}

The associated component-wise action reads
\begin{align}
S_{\text{min.}} & = \int{d^{3}{x}}\left\{  {\frac
12}i{\overline\lambda}{\gamma^{\mu}{\partial}_{\mu}}\lambda-\frac14 F_{\mu\nu}%
F^{\mu\nu}+\right. \nonumber\\
&  - A_{+}^{\ast}\Box A_{+} - A_{-}^{\ast}\Box A_{-} - \phi_{+}^{\ast}\Box \phi_{+} - \phi_{-}^{\ast}\Box \phi_{-} + i \overline\psi_{+}
\gamma^{\mu}\partial_{\mu}\psi_{+} + i \overline\psi_{-} \gamma^{\mu}%
\partial_{\mu}\psi_{-} + i \overline\omega_{+}
\gamma^{\mu}\partial_{\mu}\omega_{+} + i \overline\omega_{-}
\gamma^{\mu}\partial_{\mu}\omega_{-} \nonumber \\ & + F_{+}^{\ast}F_{+} + F_{-}^{\ast}F_{-} + S_{+}^{\ast}S_{+} + S_{-}^{\ast}S_{-} \nonumber\\
&  - q A_{\mu}\left( \overline\psi_{+}\gamma^{\mu}\psi_{+} -\overline\psi
_{-}\gamma^{\mu}\psi_{-} + \overline\omega_{+}\gamma^{\mu}\omega_{+} - \overline\omega_{-}\gamma^{\mu}\omega_{-} + iA_{+}^{\ast}\partial^{\mu}A_{+} - iA_{-}^{\ast
}\partial^{\mu}A_{-} + i\phi_{+}^{\ast}\partial^{\mu}\phi_{+} - i\phi_{-}^{\ast}\partial^{\mu}\phi_{-} \right. \nonumber \\  & \left. - iA_{+}\partial^{\mu}A_{+}^{\ast} + iA_{-}\partial^{\mu}A_{-}^{\ast} - i\phi_{+}\partial^{\mu}\phi_{+}^{\ast} + i\phi_{-}\partial^{\mu}\phi_{-}^{\ast} \right) \nonumber \\
& - iq \biggr(A_{+}\overline\psi_{+}\lambda - A_{-}\overline\psi_{-}\lambda + \phi_{+}\overline\omega_{+}\lambda - \phi_{-}\overline\omega_{-}\lambda  - A_{+}^{\ast}\overline\lambda\psi_{+} + A_{-}^{\ast}\overline\lambda\psi_{-} - \phi_{+}^{\ast} \overline\lambda\omega_{+} + \phi_{-}^{\ast} \overline\lambda\omega_{-}
\biggr) \nonumber \\ & + q^2 A_{\mu}A^{\mu}\biggr(A_{+}^{\ast}A_{+} + A_{-}^{\ast}A_{-} + \phi_{+}^{\ast}\phi_{+} + \phi_{-}^{\ast}\phi_{-} \biggr) \nonumber\\  & \left. {} - \frac{h}{2}\biggr(\phi_+^2\overline\psi_+\psi_- + \phi_+^{*2}\overline\psi_-\psi_+
- \phi_-^2\overline\psi_-\psi_+ - \phi_-^{*2}\overline\psi_+\psi_- \biggr) \, + \, \cdot\,\cdot\,\cdot\right\}\,  \nonumber
~
\end{align}
where the last line stands for the quartic interactions, as we focus on the Yukawa terms that reproduce Jackiw-Pi's chiral gauge model.
\vspace{3cm}

\section{$N$=2-Supersymmetric Generalisation}
As we pointed out in the beginning of the previous section, in order to conciliate the supersymmetric framework and the requirement of fermion number symmetry one should choose one of the following alternative (but not necessarily exclusive of the other) routes: either one incorporates the global phase transformation in the superspace structure or one enlarges the functional space, bringing extra superfields to the analysis. The latter strategy has been adopted at the $N$=1-level, as shown in the previous section. We now set up to investigate whether the former option is indeed of a feasible nature. First of all, one should notice that a global phase transformation acting on the fermionic coordinates of superpace requires a complex structure to make sense. In other words, a phase variation could not occur to real (or intrinsically real, with a real content) $\theta$ coordinates. Consequently, $N$=1-D=3 superspace does not exhibit sufficient structure to absorb the phase shift, as the $\theta$'s are Majorana fermions. As a matter of fact, one is moved to enhance the susy framework by constructing an $N$=2 theoretical model. The Grassmannian parameterisation of the $N$=2-D=3-superspace is composed of complex Dirac coordinates, upon which it is possible to define a phase transformation.

At this point we would stress the fact that performing an $N$=2 extension to a gauge theory that encompasses charged scalars is worthwhile by itself, as one provides the model with the proper framework to explore vortex solutions.  

As an algorithm to obtain an $N$=2-D=2+1 gauge theory, one may build an $N$=1-D=3+1 father model and take advantage of the different dimensionalities associated to spinorial representations of $SO(1,3)$ and $SO(1,2)$ to reduce the original $N$=1-model in the higher dimension to an $N$=2-model in the target, lower dimension. We shall proceed to the dimensional reduction at the level of superspace/superfields~\cite{teseAlv}, writing the integration measure and the superfield expansions in a four-component representation-invariant parameterisation. We then choose a diagonal Majorana representation and split each four-component D=1+3 Majorana spinor into two $SO(1,2)$-independent two-component Majorana spinors. This fermionic content is then re-written in terms of pairs of two hermitian conjugation related Dirac spinors. Also, one takes $\partial_{3}\;(\forall \; field) =  0 $ to be valid. The departure four-component $N$=1-D=4 superspace coordinates and superfields read:
\begin{eqnarray}
{\Theta} \equiv \left( 
\begin{array}{c}
\theta_a \\ \bar{\theta}^{\dot{a}}
\end{array}
\right) \label{spin1}\;,\;\;\; & {\cal W} \equiv \left(
\begin{array}{c}
W_a \\ \bar{W}^{\dot{a}}
\end{array}
\right)  \; ,
\end{eqnarray}
$\cal V $ , $\Phi$, $\overline{\Phi}$,
where the following susy-covariant constraints and equalities hold: $\cal V \, = \, {\cal V}^{\dagger} $,  $W_{a}\, = \, -\frac{1}{4}\bar{D}^{2}D_{a}\cal V$,
$\bar{W}^{\dot{a}}\, = \, -\frac{1}{4}{D}^{2}\bar{D}^{\dot{a}}\cal V$,  $\bar{D}_{\dot{a}}\Phi\, = \, 0$,  ${D}_{a}\bar{\Phi}\, = \, 0$, with
\begin{eqnarray}
D_{a} & = & \frac{\partial}{\partial\theta^{a}} \, -i \sigma^{\hat\mu}_{a\,\dot{a}}\overline{\theta}^{\dot{a}}\partial_{\hat\mu} \nonumber \\
\overline{D}_{\dot{a}} & = & -\frac{\partial}{\partial{\overline{\theta}}^{\dot{a}}} \, +i {\theta}^{a}{\overline{\sigma}}^{\hat\mu}_{a\dot{a}}\partial_{\hat\mu}\, , \nonumber  
\end{eqnarray}
where, from now on, the indices like $\hat\mu$ stand for a D=4 range, i.e., they range from 0 to 3, while indices like $\mu$ still stand for a range from 0 to 2. The $\sigma^{\hat\mu}$ matrices are just the three Pauli matrices plus a temporal component $\sigma^0$ which is the 2$\times$2 identity matrix (correspondingly, ${\bar{\sigma}}^{\hat\mu} = (1, - \sigma^{i})$).

The resulting expansions, already in four-component representation-invariant parameterisation, are  
\begin{eqnarray} {\cal V} & = & C + \bar{\Theta} B + \bar{\Theta}\Theta E - i \bar\Theta
{\Gamma}_{5}
\Theta F -
\frac{1}{2} \bar\Theta {\Gamma}_{5} {\Gamma}^{\hat{\mu}} \Theta {A}_{\hat{\mu}} +  \nonumber \\
 & & + (\bar\Theta \Theta) \bar\Theta \left(\Lambda - \frac{i}{2}{\Gamma}^{\hat{\mu}}
\partial_{\hat{\mu}} B \right) + \frac{1}{2} (\bar\Theta \Theta)^2 \left(\Delta -
\frac{1}{4}\Box C\right) \label{Vexpans}
\end{eqnarray}
\begin{eqnarray}
 {\Phi} & = &
\varphi + (\bar\Theta \Gamma_L X) +(\bar \Theta \Gamma_L \Theta)S
- \frac{i}{2} (\bar \Theta {\Gamma}^{\hat{\mu}} \Gamma_{5} \Theta)
\partial_{\hat{\mu}} \varphi  + \nonumber \\ & & -\frac{i}{2}
(\bar \Theta {\Gamma}^{\hat{\mu}} {\Gamma}_{5} \Theta)(\bar \Theta
{\Gamma}_{L}
\partial_{\hat{\mu}} X) - \frac{1}{8} (\bar \Theta \Theta)^2 \Box
\varphi  \label{phiexpans} \end{eqnarray}

\begin{eqnarray}
{\cal W}_{\alpha} & = & \Lambda_{\alpha} + \left[
2\delta_{\alpha\beta}\Delta \; + \; \frac{1}{2}{\left(
\frac{i}{2}\left[ \Gamma^{\hat{\mu}} , \Gamma^{\hat{\nu}}\right]
\right)}_{\alpha\beta} F_{{\hat{\mu}}{\hat{\nu}}}
\right]\Theta_{\beta} \; + \nonumber
\\&& +
\frac{i}{2}(\Gamma_{5})_{\alpha\beta}(\partial_{\hat{\mu}}\Lambda_{\beta})\left(
\bar{\Theta}\Gamma^{\hat{\mu}}\Gamma_{5}\Theta\right) \; + \;
\frac{i}{2}\Gamma^{\hat{\mu}}_{\alpha\beta}(\partial_{\hat{\mu}}\Lambda_{\beta})
(\bar{\Theta}\Theta) \; + \nonumber \\
&& +
\frac{i}{2}{(\Gamma_{5}\Gamma^{\hat{\mu}})}_{\alpha\beta}(\partial_{\hat{\mu}}\Lambda_{\beta})
(\bar{\Theta}\Gamma_{5}\Theta)  +
\frac{i}{2}\Gamma^{\hat{\mu}}_{\alpha\beta}(\partial_{\hat{\mu}}\Delta)\Theta_{\beta}
(\bar{\Theta}\Theta) \; + \nonumber \\ && + \frac{i}{2}
(\Gamma_{5}\Gamma^{\hat{\mu}})_{\alpha\beta}(\partial_{\hat{\mu}}\Delta)\Theta_{\beta}
(\bar{\Theta}\Gamma_{5}\Theta) + \frac{i}{8} {\left(
\Gamma_{5}(\frac{i}{2}\left[
\Gamma^{\hat{\mu}},\Gamma^{\hat{\nu}}\right])\Gamma^{\hat{\rho}}\right)}_{\alpha\beta}
(\partial_{\hat{\rho}}F_{\hat{\mu}\hat{\nu}})\Theta_{\beta}(\bar{\Theta}\Theta)\;
+ \nonumber \\ && + \frac{i}{8} {\left( (\frac{i}{2}\left[
\Gamma^{\hat{\mu}},\Gamma^{\hat{\nu}}\right])\Gamma^{\hat{\rho}}\right)}_{\alpha\beta}
(\partial_{\hat{\rho}}F_{\hat{\mu}\hat{\nu}})\Theta_{\beta}(\bar{\Theta}\Gamma_{5}\Theta)\;
-\frac{1}{8}\Box\Lambda_{\alpha}{(\bar{\Theta}\Theta)}^{2} \;\; ,
\label{W}
\end{eqnarray}
\noindent where the spinorial indices $\alpha$ and $\beta$ encompass $a$ and $\dot{a}$, thus ranging from 1 to 4 and $\Gamma^{\hat\mu}$ are the usual $D=4$ Dirac gamma matrices in the Weyl representation \footnote{\begin{eqnarray}
\Gamma^{\hat\mu} \; \equiv \;
\left(\begin{array}{cc}
0 & \sigma^{\mu}_{a\dot{b}} \\
{\bar{\sigma}}^{\mu\dot{a}b} & 0 \nonumber
\end{array}\right)  \, ; & \Gamma_{5} \; \equiv \; i\, \Gamma^{0}\,\Gamma^{1}\,\Gamma^{2}\,\Gamma^{3} \, ; & \Gamma_{L/R} \; \equiv \; \frac{1}{2}\, (1 \mp \Gamma_{5} )\, .
\end{eqnarray}}. The uncontracted index of the super-field-strength ${\cal W}_{\alpha}$ indicates a dependence on the representation, a fact that has no consequence as the only insertion of $W_{a}$ in the (super-)Lagrangian happens through the contracted square $W^{a}W_{a}$. In 4-component Majorana representation, the superspace coordinates, the spinorial superfield $\cal W$ and the spinorial component fields read
\begin{eqnarray}
B^{\mbox{\tiny Weyl}} \equiv \left(\begin{array}{c} b_a \\
\bar{b}^{\dot{a}}
\end{array}
\right) \rightarrow B^{\mbox{\tiny Major.}} \equiv \left(\begin{array}{c} b \\
d
\end{array}\right) \; ;\; & \Lambda^{\mbox{\tiny Weyl}} \equiv \left(\begin{array}{c} \lambda_a \\
\bar{\lambda}^{\dot{a}}\end{array} \right) \rightarrow
{\Lambda}^{\mbox{\tiny Major.}} \equiv
\left(\begin{array}{c} \lambda \\
\eta
\end{array}\right) \nonumber \end{eqnarray}

\begin{eqnarray}
 X^{\mbox{\tiny Weyl}} \equiv \left(\begin{array}{c} \chi_a \\
\bar{\chi}^{\dot{a}}\end{array} \right) &\rightarrow & X^{\mbox{\tiny
Major.}} \equiv
\left(\begin{array}{c} \xi \\
\omega
\end{array}\right) \nonumber \end{eqnarray}

\begin{eqnarray}
\Theta^{\mbox{\tiny Weyl}} \equiv \left(
\begin{array}{c}
\theta_{a} \\ \bar{\theta}^{\dot{a}}
\end{array}
\right) \rightarrow {\Theta}^{\mbox{\tiny Major.}} \equiv
\left(\begin{array}{c} \theta \\
\tau
\end{array}\right)\; ;\; &  {\cal W}^{\mbox{\tiny Weyl}} \equiv \left(
\begin{array}{c}
W_a \\ \bar{W}^{\dot{a}}
\end{array}
\right) \rightarrow {\cal W}^{\mbox{\tiny Major.}} \equiv \left(
\begin{array}{c}
W \\ Y
\end{array}
\right)\;\;.
\end{eqnarray}

As the dimensional reduction is performed, each 4-component Majorana spinor splits into a pair of independent two-component Majorana spinors. One can also describe the fermionic degrees of freedom by means of pairs of complex Dirac spinors, built as pairs of hermitian conjugation-related linear combinations of the Majorana spinors. For instance, one defines Dirac superfields
\begin{equation}
{\cal W}_{\pm}\equiv W \pm i Y 
\end{equation}
and Dirac fermionic component fields 
\begin{equation}
\Lambda_{\pm}\equiv \lambda \pm i \eta .
\end{equation}
This is also valid for the superspace integration measure:
\begin{eqnarray}
d^{2}\theta\; d^{2}\bar{\theta} & = & 
\frac{1}{2}(d\bar{\Theta}d\Theta)^{2} \xrightarrow{\mbox{\tiny{Major.}}}\; \frac{1}{2}(-2d\bar{\theta}d\theta
d\bar{\tau}d\tau) \nonumber \\
& = &  - d\bar{\theta}d\theta d\bar{\tau}d\tau\; , \nonumber
\end{eqnarray}
the last expression carried out through Majorana factors, while the chiral measure, $d^{2}\theta $, strongly suggests the Dirac parameterisation:
\begin{eqnarray}
d^{2}\theta & = & d\bar{\Theta}\Gamma_{L}d\Theta
\xrightarrow{\mbox{\tiny{Major.}}}\;
\frac{1}{2}(d\bar{\theta} + i d\bar{\tau})(d\theta + i d\tau) \equiv \frac{1}{2}d\bar{\theta}_{-}d\theta_{+}\, \
\label{N2chiralmeasure}
\end{eqnarray}
with the definition of complex Dirac superspace coordinates, $\theta_{\pm}\equiv \theta \pm i \tau $, being a particular case of a general recipe for 3D Dirac complex spinors, namely, $\psi_{\pm}  \equiv  \psi_{1} \pm i\psi_{2}$, where $\psi_{1}$ and $\psi_{2}$ are 3D Majorana spinors originated from the same 4-dimensional spinor. The three-dimensional superfield expansions read:  
\begin{eqnarray}
{\cal V}_{3D} & = & C + \bar{\theta} b + \bar{\theta} \theta (E + \frac{1}{2} N) + \nonumber \\
& & - \bar{\tau} \left[ d + (2F - i \slash \!\!\!\! A)\theta +
(\eta + \frac{i}{2}
\; \slash \!\!\! \partial d )\bar{\theta} \theta \right] + \nonumber \\
& & - \bar{\tau} \tau \left[ (E - \frac{1}{2} N) + \bar{\theta}
(\lambda - \frac{i}{2} \; \slash \!\!\! \partial b) + \bar{\theta}
\theta ({\Delta} - \frac{1}{4} \Box C) \right] \; =
\nonumber \\
& = & C + \frac{1}{2} (\bar{\theta}_{-} {\cal B}_{+} + \bar{\theta}_{+}
{\cal B}_{-}) + \frac{1}{2} (\bar{\theta}_{-} \theta_{+} H^* +
\bar{\theta}_{+} \theta_{-} H) + \frac{1}{2} \bar{\theta}_{-}\slash
\!\!\!\!A \theta_{-} + \frac{1}{2} \bar{\theta}_{-} \theta_{-} N + \nonumber
\\
&   & - \frac{1}{2} \bar{\theta}_{+} \left( \Lambda_{+} - \frac{i}{2}
\slash \!\!\! \partial {\cal B}_{-}\right) (\bar{\theta}_{+}
\theta_{+}) - \frac{1}{2} \bar{\theta}_{-} \left(\Lambda_{-} - \frac{i}{2}
\slash \!\!\!
\partial {\cal B}_{+}\right)
(\bar{\theta}_{-} \theta_{-}) + \nonumber \\
 &  & - \frac{1}{2} (\bar{\theta}_{-} \theta_{-})^2 \left({\Delta} - \frac{1}{4} \Box C \right) ,
\\  & & ~ \nonumber\\
\Phi_{3D} & = & \varphi +
\frac{1}{2}{\overline{\theta}}\underbrace{(\chi +
i\omega)}_{\equiv X_{+}} + \frac{1}{2}(\bar{\theta}\theta) S \; +
\nonumber \\
&& + \bar{\tau} \left[ \frac{i}{2}(\chi + i \omega) + (i S +
\partial\!\!\!\slash\varphi ) \theta -
(\frac{1}{4}\gamma^{\mu}\partial_{\mu}(\chi + i
\omega))\bar{\theta}\theta\right] \; + \nonumber \\
&& + \bar{\tau}\tau \left[ -\frac{1}{2} S +
\bar{\theta}(\frac{i}{4}\gamma^{\mu}\partial_{\mu}(\chi + i
\omega)) + (\frac{1}{4}\Box \varphi) \bar{\theta}\theta\right] \;
= \nonumber \\
&& = \;
e^{(-\frac{i}{2}\bar{\theta}_{-}\gamma^{\mu}\theta_{-}\partial_{\mu})}\left(
\varphi + \frac{1}{2}\bar{\theta}_{-}X_{+} +
\frac{1}{2}\bar{\theta}_{-}\theta_{+} S\right) \; , \nonumber \\
\overline{\Phi}_{3D} & = & \varphi^{\ast} +
\frac{1}{2}\underbrace{(\bar\chi -
i\bar\omega)}_{\equiv \bar{X}_{+}}{\theta} + \frac{1}{2}(\bar{\theta}\theta) S^{\ast} \; +
\nonumber \\
&& + \left[ \frac{-i}{2}(\bar\chi + i \bar\omega) + \bar\theta (-i S^{\ast} +
\partial\!\!\!\slash\varphi^{\ast} ) -
(\frac{1}{4}\partial_{\mu}(\bar\chi + i
\bar\omega)\gamma^{\mu})\bar{\theta}\theta\right]\tau \; + \nonumber \\
&& + \bar{\tau}\tau \left[ -\frac{1}{2} S^{\ast} +
(\frac{-i}{4}\partial_{\mu}(\bar\chi + i
\bar\omega)\gamma^{\mu}){\theta} + (\frac{1}{4}\Box \varphi^{\ast}) \bar{\theta}\theta\right] \;
= \nonumber \\
&& = \;
e^{(-\frac{i}{2}\bar{\theta}_{+}\gamma^{\mu}\theta_{+}\partial_{\mu})}\left(
\varphi^{\ast} + \frac{1}{2}\bar{\theta}_{+}X_{-} +
\frac{1}{2}\bar{\theta}_{+}\theta_{-} S^{\ast}\right) \; , \nonumber
\end{eqnarray}
where, as usual, a ``slash" means contraction with $\gamma_\mu$: $\slash \!\!\!\! A=\gamma_\mu A^\mu$, and so on. The last two expansions render evident a kind of 3D chirality, a descent of the original four-dimensional genuine chiral nature of the scalar superfield $\Phi $. The room for this planar chirality is provided by the extended superspace. At this point, one should notice that if the Dirac component-fields of ($\Phi$, $\overline{\Phi}$), $X_{+}$ and $X_{-}$, are meant to compose the 3D four-component Dirac spinor describing the electron in graphene, they should have opposite chiral gauge-charges, which is obviously true after the reciprocal charge conjugation map, and the same global phase transformation. If a global phase symmetry would be incorporated to the $N$=2-superspace coordinates structure, the complex spinors $\theta_{+}$ and $\theta_{-}$ should transform with opposite phase shifts ($\theta_{+} = {\theta_{-}}^{c}$, where the superscript $c$ means charge-conjugation)~\footnote{For the sake of completeness, we state that the 3D susy-covariant derivatives are obtained from the four-dimensional ones by means of 
$ D_{4}^{\mbox{\tiny Weyl}}\equiv \left( \begin{array}{c} {D}_{a} \\
\bar{D}^{\dot{a}} \end{array} \right)
\xrightarrow{\mbox{\tiny{Major.}}}\;\; 
D_{4}^{\mbox{\tiny Major.}}\equiv \left( \begin{array}{c} {D}_{\theta} \\
{D}_{\tau} \end{array} \right)
\rightarrow 
D_{\pm} \equiv D_{\theta} \mp i D_{\tau}\; $. 
}. We have already presented all the ingredients that one would collect to propose an $N$=2-D=3 generalisation of Jackiw-Pi's model. However, before we put forward an action functional, some remarks are mandatory: 

i) one cannot make superspace coordinates rotate globally in the same way, for they are either independent Majorana (real) parameters ($N$=1 setup) or charge conjugation (which conveys a complex conjugation)-related Dirac fermions. In the former case, no phase transformation is allowed; in the latter scenario, the coordinates must rotate with opposite phase signs. As the two independent Dirac $\theta$'s, $\theta_{+}$ and $\theta_{-}$, accompany differently gauge-charged Dirac fermions (say, $X_{+}$ and $X_{-}$ respectively) that should rotate globally in the very same way, the Grassmannian $N$=2 coordinate structure transformation must be compensated by an {\it unequal} functional global phase variation of the superfields, and the evenly charged Dirac fermions that compose the graphene electron spinor should come from {\it different} superfields; this prescription matches exactly an R-parity-like $N$=2-D=3 transformation with a particular choice of r-character attribution to each superfield; 

ii) although one manages to realise the global phase symmetry as a consequence of the extended $N$=2-superpace structure, the project of avoiding extra supermultiplets remains unfulfilled. There is a need for {\it another gauge-charged superfield}, $\Psi$, for two reasons: first, as already mentioned, if R-parity symmetry is to be defined in this theory, the two two-component Dirac fermions that span the electron spinor must come from different superfields; as for the second motivation, the spinors in $\Phi$ {\it do not fulfill the dimensionally-doubled 2+1-fermionic representation} that describe the electrons in graphene. As a matter of fact, $X_{+}$ and $X_{-}$ contain the fermionic degrees of freedom associated to a pair of 3D Majorana spinors, namely, $\chi$ and $\omega$, which match the content of {\it a single} 3D Dirac fermion. The electron in graphene demands a pair of 3D Dirac fermions to stand for its description. Thus, one should consider another matter superfield, say, $\Psi $, obeying the same susy-covariant constraints that define $\Phi$. We would like to stress that this is not an enlargement of functional space, but just an accomodation of the electronic degrees of freedom in the $N$=2 framework; 

iii) along with the completeness of fermionic degrees of freedom, one faces again the demand to establish the supermultiplet that hosts the invariant - with respect to the global phase - scalar field that enters the Jackiw-Pi's Yukawa interaction. We shall show that, w.r.t. the R-parity construction, one can easily assign global phase invariance to a scalar component that belongs either to the ``original" superfield $\Phi$ or to the ``extra" multiplet $\Psi$ (as we have just commented on, the set ($\Phi$,$\Psi$) is minimal). Nevertheless, as R-parity is a symmetry inherited from an $N$=1-D=4 setup, well defined upon chiral or anti-chiral superpotentials, the construction of an R-parity-like invariance in our $D=3$ model requires the $N$=2 analog of defined chirality, {\it excluding} superpotentials that present mixed (w.r.t. complex conjugation) terms of the kind $\Phi\Psi{\overline{\Psi}}^{2}$. On the other hand, as one deals with gauge-invariant operators, a scalar field having a graphene electron fermion component as a susy-partner - so sharing the same gauge-charge, should be brought into the Lagrangian by a compensating complex-conjugated superfield, thus defining a term that breaks the R-parity requirement for well-defined chirality. As a matter of fact, one has to consider an extra scalar superfield (also of the same susy-tensorial kind of $\Phi$), say, $\Omega$, that exhibits opposite gauge-charge with respect to $\Phi$ and $\Psi $. 

We now present the $N$=1-D=4 father model, and in the sequel the descendant $N$=2-D=3 action functional that extends Jackiw-Pi's model is depicted:
\begin{eqnarray}
{\cal S}_{4D,N=1}&=&\int d^4x\, d^{2}\theta
\;\;(\frac{-1}{16})W^{a}W_{a}  \;\; + \; h.c.  \nonumber \\ & +
& \int d^4 x \, d^{2}\theta d^{2}\overline{\theta}
\left\{ \frac
{1}{16}\left(\overline{\Phi }e^{2h{\cal V}}\Phi + \overline{\Psi }e^{2h{\cal V}}\Psi + \overline{\Omega}\, e^{-2h{\cal V}}\Omega \right)  \right\} \, + \nonumber \\ & + & \int d^4x\, d^{2}\theta
\;\; \left( 4g\,\Phi\,\Psi\, \Omega^{2} \right)\;\; + \;\; \int d^4x\, d^{2}\overline{\theta}
\;\; \left( 4g\,\overline{\Phi}\,\overline{\Psi}\, \overline{\Omega}^{2} \right)
 \; ,\label{fathermodel}
\end{eqnarray}
where the super-fieldstrength, $W_{a}$, is defined as $W_{a} = -\frac{1}{4}{\bar{D}}^{2} D_{a}\cal{V}$, and the vector-superfield $\cal{V}$ reads as presented in eq. (\ref{Vexpans}). Also, the superfields $\Phi$, $\Psi$ and $\Omega$ share the same susy-covariant constraint (chiral superfields) with the multiplet expanded in eq. (\ref{phiexpans}). The dimensionally reduced action now follows:

\begin{eqnarray}
{\cal S}_{3D,N=2}&=&\int d^3x\, d\overline{\theta}_{-}d\theta_{+}
\;\;(\frac{-1}{32})\overline{\cal W}_{-}{\cal W}_{+}  \;\; +  \nonumber \\ & +
& \int d^3 x \,d\overline{\theta}d\theta d\overline{\tau}d\tau
\left\{ - \frac
{1}{16}\left(\overline{\Phi }e^{2h{\cal V}}\Phi + \overline{\Psi }e^{2h{\cal V}}\Psi + \overline{\Omega}\, e^{-2h{\cal V}}\Omega \right)  \right\} \, + \nonumber \\ & + & \int d^3x\, d\overline{\theta}_{-}d\theta_{+}
\;\; \left( 2g\,\Phi\,\Psi\, \Omega^{2} \right)\;\; + \;\; \int d^3x\, d\overline{\theta}_{+}d\theta_{-}
\;\; \left( 2g\,\overline{\Phi}\,\overline{\Psi}\, \overline{\Omega}^{2} \right)
 \; .\label{eq2}
\end{eqnarray}
The superfields $\Psi$ and $\Omega$ of the latter functional hold the following expansions:
\begin{eqnarray}
\Psi_{3D} & = & \rho +
\frac{1}{2}{\overline{\theta}}\underbrace{(\xi +
i\beta)}_{\equiv \psi_{+}} + \frac{1}{2}(\bar{\theta}\theta) P \; +
\nonumber \\
&& + \bar{\tau} \left[ \frac{i}{2}(\xi + i \beta) + (i P +
\partial\!\!\!\slash\rho )\theta -
(\frac{1}{4}\gamma^{\mu}\partial_{\mu}(\xi + i
\beta))\bar{\theta}\theta\right] \; + \nonumber \\
&& + \bar{\tau}\tau \left[ -\frac{1}{2} P +
\bar{\theta}(\frac{i}{4}\gamma^{\mu}\partial_{\mu}(\xi + i
\beta)) + (\frac{1}{4}\Box \rho) \bar{\theta}\theta\right] \;
= \nonumber \\
&& = \;
e^{(-\frac{i}{2}\bar{\theta}_{-}\gamma^{\mu}\theta_{-}\partial_{\mu})}\left(
\rho + \frac{1}{2}\bar{\theta}_{-}\psi_{+} +
\frac{1}{2}\bar{\theta}_{-}\theta_{+} P\right) \; , \nonumber
\end{eqnarray}
\begin{eqnarray}
\Omega_{3D} & = & \phi +
\frac{1}{2}{\overline{\theta}}\underbrace{(\delta +
i\sigma)}_{\equiv \kappa_{+}} + \frac{1}{2}(\bar{\theta}\theta) M \; +
\nonumber \\
&& + \bar{\tau} \left[ \frac{i}{2}(\delta + i \sigma) + (i M +
\partial\!\!\!\slash\phi )\theta -
(\frac{1}{4}\gamma^{\mu}\partial_{\mu}(\delta + i
\sigma))\bar{\theta}\theta\right] \; + \nonumber \\
&& + \bar{\tau}\tau \left[ -\frac{1}{2} M +
\bar{\theta}(\frac{i}{4}\gamma^{\mu}\partial_{\mu}(\delta + i
\sigma)) + (\frac{1}{4}\Box \phi) \bar{\theta}\theta\right] \;
= \nonumber \\
&& = \;
e^{(-\frac{i}{2}\bar{\theta}_{-}\gamma^{\mu}\theta_{-}\partial_{\mu})}\left(
\phi + \frac{1}{2}\bar{\theta}_{-}\kappa_{+} +
\frac{1}{2}\bar{\theta}_{-}\theta_{+} M\right) \; , \nonumber
\end{eqnarray}
with the $\overline{\Psi}$ expansion bringing about $\rho^{\ast}$ and $\psi_{-}\equiv \xi - i \beta$, and the $\overline{\Omega}$ expansion exhibiting the fields $\phi^{\ast}$ and $\kappa_{-}\equiv \delta - i \sigma$ as components. The symmetries are:

Gauge transformation:
\begin{eqnarray}
\Phi^{\prime} & = & e^{2ih\Lambda}\Phi \nonumber \\
\Psi^{\prime} & = & e^{2ih\Lambda}\Psi \nonumber \\
\Omega^{\prime} & = & e^{-2ih\Lambda}\Omega \nonumber \\
{\cal V}^{\prime} & = & {\cal V} + i \left( {\Lambda} - \overline{\Lambda}\right) \; , \nonumber
\end{eqnarray}
where $\Lambda$ is an $N$=2 parametric superfield that obeys the same susy-constraints defining $\Phi$.

Parity, acting on the superspace coordinates:
\begin{eqnarray}
{\left(\begin{array}{c}
\theta \\
\tau
 \end{array}\right)}^{\prime} & = & \left(\begin{array}{cc}
0 & \gamma^0\gamma^2 \\
\gamma^0\gamma^2 & 0 
 \end{array}\right) \; {\left(\begin{array}{c}
\theta \\
\tau
 \end{array}\right)} \, , 
\end{eqnarray}
which implies $\theta_{+}^{P}\, = \, i \gamma^{0}\gamma^{2} \theta_{-}$ and $\overline{\theta}_{-}^{P}\, = \, -i {\overline{\theta}}_{+}\gamma^{0}\gamma^{2}$, leading to
\begin{eqnarray}
d\bar{\theta}_{-}d\theta_{+} &
\xrightarrow{\mbox{\tiny{Parity}}} &
d\bar{\theta}_{+}d\theta_{-}\, 
\nonumber
\end{eqnarray}
It is now evident that Parity relates chiral and anti-chiral sectors of $N$=2-D=1+2 superspace. Moreover, the fermionic field meant to describe the electron in graphene should be composed by two independent complex Dirac spinors, a fact that indicates that the appropriate candidate is a pair $\left( \psi_{+}, X_{-} \right)$, or equivalently a pair $\left( X_{+}, \psi_{-} \right)$ \footnote{Notice that the two pairs are {\it not} independent degrees of freedom; in that sense, we face no superfluous field variable.}. On this token, Parity should map $\psi_{+}$ onto $X_{-}$, and conversely, bring $X_{-}$ back to $\psi_{+}$. Consequently, as $\psi_{+}$ is a component-field in the expansion of $\Psi$ and $X_{-}$ plays the analogous role in the expansion of $\overline{\Phi}$ (mind the bar!!), the action of Parity at the level of superfields yields
\begin{eqnarray}
\Phi^{P} & \rightarrow &  \overline{\Psi} \nonumber \\
\Psi^{P} &\rightarrow &  \overline{\Phi} \nonumber \\
\Omega^{P} &\rightarrow &  \overline{\Omega} \nonumber \\
{\cal V}^{P} &\rightarrow & {\cal V} \, ,
\end{eqnarray} 
and the Parity role in the interplay of $N$=2 chiralities is again evident. This picture is fully consistent with the fact that the two components of the pair (of complex 3D Dirac spinors) that describe the electron in graphene should have the same global phase/fermion number transformation, while exhibiting opposite gauge-charges. In the following subsection, we present the R-parity prescription for the global phase symmetry associated to electric charge in graphene.
\subsection{The R-Parity Prescription}
We define R-parity transformation, in an $N$=1-D=4 framework, as the result of combined $\theta$ (and $\overline{\theta}$)-coordinate rotation and superfield functional phase variation. We shall take advantage of the dimensional reduction procedure described in the beginning of this Section to establish the corresponding $N$=2-D=3-$\theta_{+},\theta_{-}$-coordinate and superfields transformation. According to (\ref{phiexpans}), a D=4 term like $\theta^{a}\psi_{a}$ can be written in a representation-invariant formulation as $\overline{\Theta}\Gamma_{L}\Psi$. The variation $\theta^{a} \;\xrightarrow{\mbox{\tiny{R-Parity}}} e^{-i\alpha} \theta^{a} \;$, which implies $\theta^{a}\psi_{a} \;\xrightarrow{\mbox{\tiny{R-Parity}}} e^{-i\alpha} \theta^{a}\psi_{a} \;$, has its $N$=2-D=3 analog obtained through the diagonal Majorana representation reading of $\overline{\Theta}\Gamma_{L}\Psi$, that happens to be equal to $\frac{1}{2}\overline{\theta}_{-}\psi_{+}$. One then establishes that $\overline{\theta}_{-} \;\xrightarrow{\mbox{\tiny{R-Parity}}} e^{-i\alpha} \overline{\theta}_{-} \;$. The same conclusion comes out after the analysis of the D=4-fermionic measure R-parity variation (a consequence of its Berezinian derivative nature after the coordinates multiplication by global phase factors), as one follows (\ref{N2chiralmeasure}):
\begin{eqnarray}
d^{2}\theta \;\xrightarrow{\mbox{\tiny{R-Parity}}} e^{-2i\alpha} d^{2}\theta & ; & d^{2}\overline{\theta} \;\xrightarrow{\mbox{\tiny{R-Parity}}} e^{+2i\alpha} d^{2}\overline{\theta}\nonumber \\
d^{2}\theta \;\xrightarrow{\mbox{\tiny{Dim.Reduc.}}} \frac{1}{2} d\overline{\theta}_{-}d\theta_{+} & \xrightarrow{\mbox{\tiny{R-Parity}}} & e^{-2i\alpha} \left(\frac{1}{2} d\overline{\theta}_{-}d\theta_{+}\right) \nonumber \\
d^{2}\overline{\theta} \;\xrightarrow{\mbox{\tiny{Dim.Reduc.}}} \frac{1}{2} d\overline{\theta}_{+}d\theta_{-} & \xrightarrow{\mbox{\tiny{R-Parity}}} & e^{+2i\alpha} \left(\frac{1}{2} d\overline{\theta}_{+}d\theta_{-}\right)\nonumber \; , 
\end{eqnarray}
implying again $\overline{\theta}_{-} \;\xrightarrow{\mbox{\tiny{R-Parity}}} e^{-i\alpha} \overline{\theta}_{-}$ and $\overline{\theta}_{+} \;\xrightarrow{\mbox{\tiny{R-Parity}}} e^{+i\alpha} \overline{\theta}_{+}$ as well. For a generic ``chiral'' $N$=2-D=3-superfield $\Sigma$ (equivalent to $\Phi$, $\Psi$ and $\Omega$), the action of R-Parity transformation reads:
\[
\Sigma \xrightarrow{\mbox{\tiny{R-Parity}}} e^{2i({n}_{\Sigma})\alpha}\Sigma(e^{-i\alpha}\overline{\theta}_{-}, x),
\]
where $n_{\Sigma}$ stands for the r-character of the $\Sigma$-superfield. Consequently, the generic anti-chiral $\overline{\Sigma}$ varies according to $\overline{\Sigma} \xrightarrow{\mbox{\tiny{R-Parity}}} e^{-2i({n}_{\Sigma})\alpha}\overline{\Sigma}(e^{+i\alpha}\overline{\theta}_{+}, x)$. If we assume the supermultiplet $\Sigma$ to contain the component-fields ($s,\sigma_{+},S$), where $s(x)$ is a physical scalar field, $\sigma_{+}(x)$ is a D=3-Dirac Fermi field and $S(x)$ is a scalar auxiliary field (and, correspondingly, $\overline{\Sigma} = \overline{\Sigma}(s^{*},\sigma_{-}(x),S^{*}(x)$), the component-wise R-parity variation is as follows:
\begin{eqnarray}
s(x) & \xrightarrow{\mbox{\tiny{R-Parity}}} & e^{2i(n_{\Sigma})\alpha} s(x) \nonumber \\
\sigma_{+} (x ) & \xrightarrow{\mbox{\tiny{R-Parity}}} & e^{2i(n_{\Sigma} - 1/2)\alpha}\, \sigma_{+}(x) \label{Rparfermion}\\
S(x) & \xrightarrow{\mbox{\tiny{R-Parity}}} & e^{2i(n_{\Sigma} - 1)\alpha}\, S(x) \nonumber \, ,  
\end{eqnarray}
and the analogous expressions for the $\overline{\Sigma}$ components read:
\begin{eqnarray}
s^{*}(x) & \xrightarrow{\mbox{\tiny{R-Parity}}} & e^{-2i(n_{\Sigma})\alpha} s^{*}(x) \nonumber \\
\sigma_{-} (x ) & \xrightarrow{\mbox{\tiny{R-Parity}}} & e^{-2i(n_{\Sigma} - 1/2)\alpha}\, \sigma_{-} (x) \label{Rparfermion2}\\
S^{*}(x) & \xrightarrow{\mbox{\tiny{R-Parity}}} & e^{-2i(n_{\Sigma} - 1)\alpha}\, S^{*}(x) \nonumber \, .  
\end{eqnarray}
Now, let us consider the following particular assignment of r-character ($n_{\Sigma}$) to the superfields present in the proposed action (\ref{eq2}):

$n_{\Phi} \, = \, 1 \, ; \, n_{\Psi} \, = \, 0 \, ; \, n_{\Omega} \, = \, 0\, .$\footnote{To properly compensate the variation of the integration measure, it is a well establish result that the sum of the r-characters assigned to superfields that compose a term in the superpotential must be one.}

As a consequence, the pair of D=3-two-components Dirac fermions that describe the graphene electronic degrees of freedom, namely, $(X_{+}, \psi_{-})$, integrated out of the superfields $(\Phi, \overline{\Psi})$, transform according to

\begin{eqnarray}
X_{+} & \xrightarrow{\mbox{\tiny{R-Parity}}} & e^{2i(1-1/2)\alpha} X_{+} \, = \, e^{+i\alpha}X_{+} \nonumber \\
\psi_{-} & \xrightarrow{\mbox{\tiny{R-Parity}}} & e^{-2i(0-1/2)\alpha} X_{+} \, = \, e^{+i\alpha}\psi_{-} \nonumber \, . 
\end{eqnarray}
Thus, the components of the graphene electron spinor transform with the same global phase, allowing for the interpretation of the fermion number/electric charge symmetry as an R-parity-like superspace invariance. To complete the reasoning, one can easily verify that the scalar field meant to play a role in the Jackiw-Pi's Yukawa term happens to be invariant (neutral) under the R-parity-like (fermion number) transformation: as $\phi$ belongs to the $\Omega$ supermultiplet, and $n_{\Omega} = 0$, $\phi \xrightarrow{\mbox{\tiny{R-Parity}}} e^{2i(n_{\Omega})\alpha} \phi \, = \, \phi $. The set of symmetries is then complete\footnote{The model is also invariant under an additional Global Phase Transformation:
\begin{eqnarray}
\Phi^{\prime} & = & e^{i\alpha}\Phi \nonumber \\
\Psi^{\prime} & = & e^{-i\alpha}\Psi \nonumber \\
\Omega^{\prime} & = & \Omega \nonumber\\
{\cal V}^{\prime} & = & {\cal V} \nonumber \; .
\end{eqnarray} We could as well take this global functional variation of superfields as the prescription for the global fermion number/electric charge symmetry. We nevertheless shall keep the R-parity-like interpretation for the electric charge, as we wish to explore all the possibilities associated to the $N$=2-extended supersymmetric framework.}
, as we have been able to express the local-U(1) (axial gauge) symmetry, the discrete Parity symmetry and the R-parity-like global phase (fermion number/electric charge) invariance. 
Finally, the $N$=2-D=3 supersymmetric component-wise action reads: 
\begin{eqnarray}
{\cal S}_{3D,N=2} &=&\int d^3x\left\{ -\frac{1}{4}F_{\mu \nu
}F^{\mu \nu }+2\Delta ^2+\frac{1}{2}\partial _\mu N\partial ^\mu N
+ hv^2 \Delta + \right. \nonumber \\ &&+\frac{1}{2}\bar{\Lambda
}_{-}(i\partial \!\!\!/)\Lambda _{-}+ \left[ (D
_{+\mu} \varphi)( D_{+}^\mu \varphi) ^{*}+  (D
_{+\mu} \rho )( D_{+} ^\mu \rho) ^{*}+  
(D_{-\mu} \phi)( D_{-} ^\mu \phi) ^{*} + \right. \nonumber \\ &&  -(h^2N^2 - 2h\Delta)\,\varphi
\varphi ^{*}  -(h^2N^2 - 2h\Delta)\,\rho
\rho^{*}  -(h^2N^2 + 2h\Delta)\,\phi
\phi ^{*} +\left| S\right| ^2 +\left| P\right| ^2 +\left| M\right| ^2  \nonumber \\ &&+\frac{1}{4}(hN)\,\bar{X}_{+}X_{+} +\frac{1}{4}(hN)\,\bar{\psi}_{+}{\psi}_{+} +\frac{1}{4}(hN)\,\bar{\kappa}_{+}{\kappa}_{+}\nonumber \\ && +\frac{i}{8}(\bar{X}_{-}D
\!\!\!\!/\,_{-}X_{-}+\bar{X}_{+}D \!\!\!\!/\,_{+} X_{+}) +\frac{i}{8}(\bar{\psi}_{-}D
\!\!\!\!/\,_{-}{\psi}_{-}+\bar{\psi}_{+}D \!\!\!\!/\,_{+} {\psi}_{+}) +\frac{i}{8}(\bar{\kappa}_{-}D
\!\!\!\!/\,_{+}\kappa_{-}+\bar{\kappa}_{+}D \!\!\!\!/\,_{-} \kappa_{+})
\nonumber \\
&& \left.  -\frac{h}{2}\left[ (\varphi \bar{\Lambda }_{+}X_{-}+\varphi
^{*}\bar{X}_{-}\Lambda_{+}) +(\rho \bar{\Lambda}_{+}{\psi}_{-}+\rho
^{*}\bar{\psi}_{-}\Lambda_{+}) -(\phi \bar{\Lambda}_{+}\kappa_{-}+\phi
^{*}\bar{\kappa}_{-}\Lambda_{+})\right]   \right] \nonumber \\
&& + g  \left[ \rho\varphi(\bar{\kappa}_{-}\kappa_{+}) + 2\phi\varphi(\bar{\psi}_{-}\kappa_{+}) + 2\rho\phi(\bar{X}_{-}\kappa_{+}) + \underbrace{
{\phi}^{2}(\bar{\psi}_{-}X_{+})}_{\mbox{J-P's} \; \mbox{Yukawa}\; \mbox{int.}}  + \right. \nonumber \\
&& \left. \left. -4 (2M\phi\varphi\rho + P\varphi{\phi}^{2} +  S\rho{\phi}^{2} )+ h.c. \right] \right\}, \label{demanda}
\end{eqnarray}
where $D_{+\mu} (\varphi,\rho) \equiv (\partial_{\mu} + ih A_{\mu})(\varphi,\rho) $, and $D_{-\mu} \phi \equiv (\partial_{\mu} - ih A_{\mu})\phi $. One should note the insertion of a gauge symmetric Fayet-Iliopoulos term~\cite{FI}\footnote{Such a term does not affect the $N$=2-symmetry of the action, as it $N$=2-transforms as a total derivative.}, $ hv^2 \Delta$, meant to allow for topologically non-trivial vortex solutions.

One can now recognise, among the set of terms spanned by the quartic interaction $g\Phi\Psi\Omega^{2}$  and its hermitian conjugate, the operators $g{\overline{\psi}_{-}} X_{+} \phi^{2} + g\overline{\psi}_{+} X_{-}{\phi^{\ast}}^{2}$, endowed with all the properties and transformation rules that define the Yukawa terms in the Jackiw-Pi's model, Eq.~(\ref{LHCM2}).

\section{Bogomol'nyi equations: the minimum energy vortex configuration ansatz}

An extended supersymmetry algebra allows for a non-trivial minimum (bound) for the energy functional whenever half of the supersymmetries (one of the two we work upon in an $N$=2 context) are taken to have a null transformation effect onto the fields~\cite{OliveWitten, HS1, HS2, HS3}. Such a bound happens to be associated to the central charge of the extended susy algebra. Also, the bound conveys topological information, as one can show that its value is proportional to the magnetic flux, a fact that leads to a direct relation between the central and the topological charges. The null variation of the fermionic fields with respect to one of the supersymmetries gives rise to first-order differential equations obeyed by the bosonic fields at the bound, then said to be saturated. So, a minimum energy configuration satisfies the following set of Bogomol'nyi equations\footnote{We assume the Wess-Zumino gauge to be fixed. Such a non-susy-covariant gauge-fixing choice imposes a redefinition of susy transformation group and algebra, and the resulting Wess-Zumino gauge preserving graded algebra mixes the original susy and U(1) transformations, turning the space-time derivative representation of the translation operator, $P_{\mu} = i\partial_{\mu}$, into a U(1)-gauge covariant extension, where the translation is represented by $D_{\mu} = \partial_{\mu} + ih A_{\mu} $.}:

\begin{eqnarray}
&& B \pm 2 \Delta =0 \nonumber
\\ & & E_{i} \pm \partial_{i} N=0  \nonumber \\ & & \partial_{0}N = 0 \nonumber \\
& & D_{+0}\varphi \pm ihN\varphi =0  \nonumber \\ & & \left( D_{+1} \mp i
D_{+2}\right) \varphi =0  \label{selfdual1} \\
&& S = 0 \label{auxiliar1} \; ; \\& & D_{+0}\rho \pm
ihN\rho =0  \nonumber \\ & & \left( D_{+1} \mp i
D_{+2}\right) \rho =0  \label{selfdual2} \\ && P = 0 \label{auxiliar2} \; ; \\
& & D_{-0}\phi \mp ihN\phi =0  \nonumber \\ & & \left( D_{-1} \mp i
D_{-2}\right) \phi =0  \label{selfdual3} \\
&& M = 0 \label{auxiliar3} \; . 
\end{eqnarray}

As one proceeds to writing the set of Bogomol'nyi equations in terms of bosonic physical fields, the auxiliary-fields $\Delta$, $S$, $P$ and $M$ must be taken on-shell. The corresponding algebraic field-equations read:
\begin{eqnarray}
\Delta & = & - \frac{h}{2}\left(  |\varphi|^{2} + |\rho|^{2} - |\phi|^{2} + \frac{v^{2}}{2}  \right) \; ; \nonumber \\
S & = & 4g \rho^{*}{\phi^{*}}^{2} \; ; \nonumber \\
P & = & 4g \varphi^{*}{\phi^{*}}^{2} \; ; \nonumber \\
M & = & 8g \varphi{*}\rho^{*}\phi^{*} \; . \label{onshellaux}
\end{eqnarray}

The Bogomol'nyi equations (\ref{auxiliar1}), (\ref{auxiliar2}), (\ref{auxiliar3}) then imply that $\rho^{*}{\phi^{*}}^{2} = 0$, $\varphi^{*}{\phi^{*}}^{2} = 0$ and $\varphi^{*}\rho^{*}\phi^{*} = 0$ simultaneously for a bound-saturating field configuration. Such a condition may be fulfilled in two different ways: either one takes $\varphi = \rho = 0$, with a remaining non-trivial $\phi$ or, alternatively, one takes $\phi = 0$, and leaves $\varphi$ and $\rho$ so far undetermined. At this point, one should account for the potential read from the action functional (\ref{demanda}): $U\, = \, 2 \Delta^{2} + {|S|}^{2} + {|P|}^{2}+ {|M|}^{2} + h^{2}N^{2}({|\varphi|}^{2} + {|\rho|}^{2} + {|\phi|}^{2})$. As a matter of fact, only the first route, keeping the $\phi$-field non-trivial and assuming null configurations for the scalars $\varphi$, $\rho$ and $N$ matches the minimum of the potential, at zero value\footnote{As far as the scalar-field $\phi$ is concerned, the potential does not allow for a minimum non-topological configuration.}. So motivated, we  
consider the $\phi$-field as a candidate to provide the non-perturbative spectrum with a vortex excitation associated to the coupling  $\phi \leftrightarrow \vec{A}$. As a byproduct, saturating the bound this way, with vanishing $\varphi$ and $\rho$, reduces the field-content of the model for a minimum energy configuration, and no need remains, at the bound, for a phenomenological interpretation of the graphene electronic degrees of freedom scalar partners\footnote{One should note that the sign of the Fayet-Iliopoulos term has been so chosen that a topologically non-trivial minimum energy configuration for the $\phi$-field is allowed. Had we considered the second possibility, with non-vanishing $\varphi$ and $\rho$-fileds, the sign of the F.-I. term should accordingly be changed.}. Additionally, one should note that the $\phi$-field is regarded as the scalar field-component of the Jackiw-Pi's Yukawa interaction term\footnote{As a matter of fact, $-\phi^2$ has a complex conjugation relation with J.-P's scalar $\varphi$, as one compares Eq.~(\ref{demanda}) to Eq.~(\ref{LHCM2}).}. Therefore, a vortex configuration ansatz defined upon $\phi$ keeps the track of a Jackiw-Pi's extended theory proposal.

One can rephrase the Bogomol'nyi equations, as the set of field-equations (\ref{onshellaux}) prevail and the choice to keep the $\phi$-field is taken:
\begin{eqnarray}
&& B \pm  \frac{h}{2}\left( + 2|\phi|^{2} - {v^{2}}  \right) =0 \label{Bphi}
\\ & & E_{i} \pm \partial_{i} N=0  \nonumber \\ & & \partial_{0}N = 0 \label{estat} \\
& & D_{-0}\phi \mp ihN\phi =0  \nonumber \\ & & \left( D_{-1} \mp i
D_{-2}\right) \phi =0  \label{selfdualphi} \\
&& M = 0 \label{auxiliarphi} \; . 
\end{eqnarray}

The equation (\ref{estat}) implies static configuration when the bound is saturated. Such a fact is easily verified by means of reimposing (\ref{estat}) everywhere in the Bogomol'nyi set of equations. Equations (\ref{Bphi}) and (\ref{selfdualphi}) are the ones to be satisfied by the following vortex configuration ansatz\footnote{These equations are the same found for the self-dual Maxwell-Higgs model \cite{de Vega}, but one should note that the present model yields a different potential.}:

\begin{eqnarray}
\phi(r, \theta) & = & v\, R(r)e^{in\theta} \; , \label{phivort} \\
{\mbox{\bf A}} & = & -\, \frac{\hat{\theta}}{hr}\left[ a(r) -
n\right] \; . \label{Avort}
\end{eqnarray}

\noindent In the {\it ansatz} above, $R(r)$ and $a(r)$ are 
real functions of $r$, and the vorticity, $n$, is an integer 
that labels the topological charge of the configuration.
The magnetic field strength and the corresponding flux are:
\begin{eqnarray}
B\; = \; -\frac{1}{hr}\frac{da}{dr} & \rightarrow & \Phi_{B} \; =
\; \frac{2\pi}{h}\left[ a(0) - a(\infty) \right]\; . \label{Bflux}
\end{eqnarray}

Bogomol'nyi equations (\ref{Bphi}) and (\ref{selfdualphi}) then read, as one adopts the vortex configuration presented above:
\begin{eqnarray}
\frac{dR}{dr} \pm
\frac{a}{r} R & = & 0 \; , \label{bogvort} \\
\frac{1}{r}\frac{da}{dr} \mp \frac{h^{2}v^{2}}{2}(2R^{2}
-1) & = & 0 \; . \label{segvort}
\end{eqnarray}

Appropriate boundary conditions, with the assumption of finite energy
(which parameterises the behavior at infinity) and of non-singular behavior 
(which controls $R(0)$ and $a(0)$), associated to 
Eqs.~(\ref{bogvort}) and (\ref{segvort}), allow one to probe, by means of numerical analysis~\cite{selfdual}, 
topological vortices in the spectrum of the present model. 
The boundary conditions for the topological configuration are:
$R(\infty)=\frac{1}{\sqrt{2}}$, $a(\infty)=0$, $nR(0)=0$, $a(0) = n$.
Consequently, the flux (\ref{Bflux}) turns out to be quantised: $\Phi_{B} =
\frac{2\pi}{h}\left[ a(0) - a(\infty) \right] = 2\pi n/h $.

The solution close to the origin assumes the form\footnote{Power series method was employed; the coefficients $c_{n}$ can be numerically determined from the asymptotic behaviour of the functions at infinity, and for the present simulation they read $c_{1} = 1.4154\times 10^{-2}$, $c_{2} = 1.4167\times 10^{-4} $ and $c_{3} = 1.0916\times 10^{-6} $.} %
\begin{align*}
R(r)  & = \,  c_{n}r^{n} + \cdots ,\\
a(r)  & =\, n-\frac{h^{2}v^{2}}{4}r^{2} + \cdots .
\end{align*}

As it takes to evaluate the functions $R(r)$ and $a(r)$ as $r$ approaches infinity, one adopts the solutions of the system of equations (\ref{bogvort}) and (\ref{segvort}) for infinitesimal variations apart from the corresponding boundary values \cite{Casana}. On this token, one searches for the radial dependence of $\epsilon_{R}(r)$ and $\delta\, a(r)$, defined as $R(r{\scriptstyle \rightarrow}\infty) = \frac{1}{\sqrt{2}} - \epsilon_{R}(r)$ and $a(r{\scriptstyle \rightarrow}\infty) = 0 + \delta\, a(r)$. It comes out from Eqs.~(\ref{bogvort}) and (\ref{segvort}) that
\begin{align*}
\epsilon_{R}(r)  & \sim r^{-1/2}e^{-hvr},\\
\delta\, a(r)  & \sim r^{1/2}e^{-hvr}.
\end{align*}

Assuming the $r$-dependence profile that prevails around the origin as well as the behavior for large $r$-values, and taking into account the boundary conditions, we have numerically solved the equations of motion.
In Fig.~1, we plot\footnote{The minimal coupling constant is taken at the value $h=1$, and for the non-trivial vacuum expectation value parameter one takes $v = 0.3$.} the results for $R(r)$ and $a(r)$, with vorticity numbers $n=1$, $2$ and
$3$.  
In Fig.~2, we show the magnetic field for the same set of parameter values adopted for Fig.~1. 
\vspace{\baselineskip}

\begin{figure}
[ht]
\begin{center}
\includegraphics[angle=0,width=0.81\linewidth]
{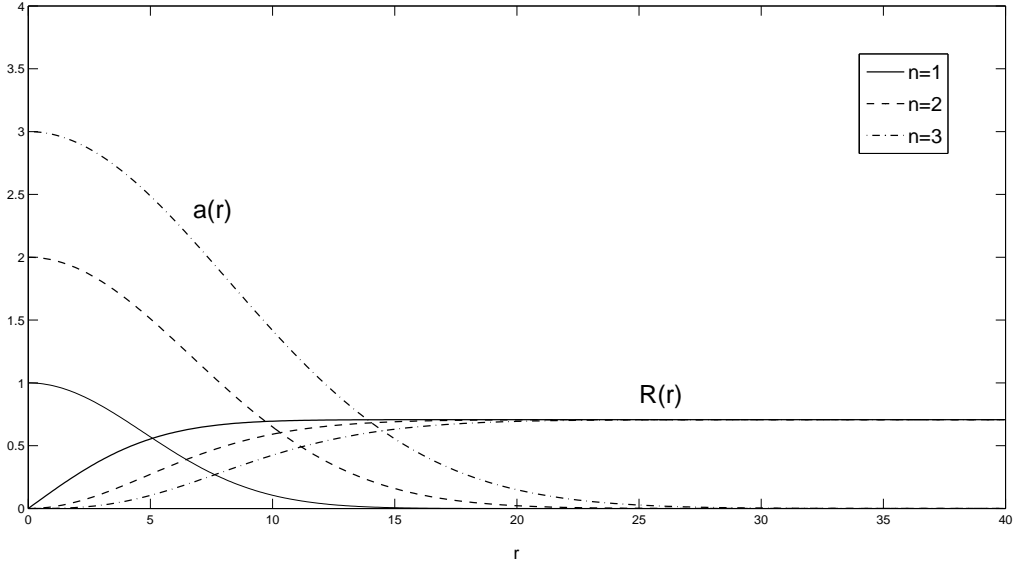}%
\end{center}
\caption{\label{fig1} \it Scalar R(r) and gauge field component-function a(r) in the topological configuration for n=1, 2, 3.}
\end{figure}

\begin{figure}
[h]
\begin{center}
\includegraphics[angle=0,width=0.81\linewidth]
{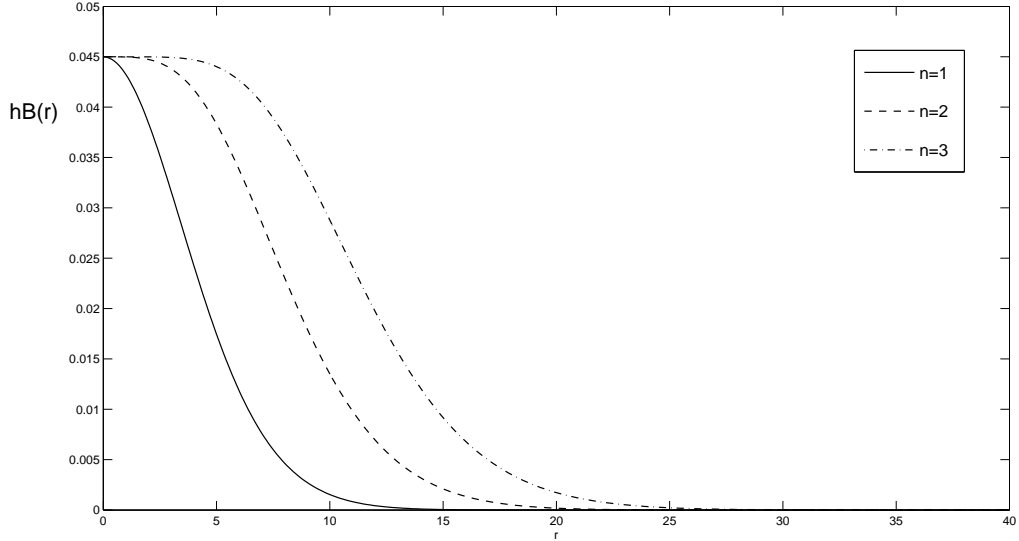}%
\end{center}
\caption{\label{fig2} \it The magnetic field B as function of r in the topological configuration for n=1, 2, 3.}
\end{figure}

\section{Conclusions and Perspectives} 
In a recent work~\cite{nosso}, we suceeded to show that the description of electronic degrees of freedom in monolayer graphene could be formulated in a supersymmetric framework. Moreover, the $N$=1 extension may be constructed in connection with the $N$=1-$\tau_{3}$-QED, which in its turn stems from a dimensional reduction of an $N$=1 gauge model built on Atiyah-Ward (2+2)-spacetime. However, such an approach failed to conciliate the fermion number symmetry (that would be associated to the electric charge) and the Yukawa term proposed by Jackiw and Pi. 

Nevertheless, a straightforward $N$=1-extension of Jackiw-Pi's model was constructed in the present work, allowing for the Yukawa interaction while conserving fermion number, at the expense of introducing an extra pair of complex scalar superfields. Concerning the phenomenology of graphene, one is left with the challenge of providing physical interpretation to the bunch of extra fields needed to span a supersymmetric generalisation. As we tried to avoid extra superfields, the quest for a sort of embedding of global phase transformation in the superspace structure led us to the formulation of a further extended $N$=2-D=1+2-model. In such a framework, although extra superfields remained essential for the model, an R-parity-like picture was shown to account for the global phase/fermion number symmetry, allowing for a quite interesting physical connection between the electric charge conservation and an inherent superspace structure invariance. 

At this point, however, we would like to stress that the $N$=2-extension has merits of its own, and this fact was explored here as a basis for the search for topologically non-trivial non-perturbative solutions. We started from the well known fact that an extended supersymmetry is the proper framework to establish the set of dynamical equations that provide the spectrum with, for instance, vortex excitations. Such equations happen to be the result of imposing triviality on half of the SUSY variations of fermionic fields. Also, the central charge in the supersymmetric algebra plays the role of the topological charge underlying soliton-like solutions. In this context, our results, in the present work, are the achievement of Bogomol'nyi equations and the corresponding numerical vortex solution, which still need to be properly compared in detail with other vortex solutions for graphene in the literature, especially those predicted by Jackiw's et al.\ model. Such a comparison will soon be reported elsewhere. 

From another perspective, the issue of spontaneous supersymmetry breaking deserves attention as the $N$=2-D=1+2 setup mimics an $N$=1-D=1+3 structure, and a full analysis of the possible minima for the corresponding potential is still lacking. The result of this analysis is crucial to define mass eigenstates and is certainly of valuable help to understand the physical r{\^o}le of the extra fields. Our efforts to elucidate these matters will also be reported elsewhere.

Finally, we would like to make some comments about Majorana fermions. There is already evidence for their experimental realisation in condensed matter systems (see, for example, Refs.~\cite{Mourik, Stanescu}), which do not include graphene. However, in the case of the latter material, theoretical predictions do exist: see, for example, Refs.~\cite{MF3, MF-Graph-like, BitanRoy}. All of those models commonly require superconductivity to be present, an exception being the work of Ref.~\cite{MF-trig}, in which the key ingredient is the explicit consideration of the so-called trigonal warping term in the Hamiltonian of graphene. In the present work, the $\lambda$-field, the supersymmetric partner of the bosonic field $A_\mu$, is a Majorana fermion ``by birth". However, in order to properly state that Majorana fermions are present in our model, it is necessary to investigate whether massless and neutral eigenvectors arise from diagonalisation of the fermionic mass matrix evaluated at scalar fields configurations that match the minimum potential condition (gauge symmetry breaking). Furthermore, if this actually occurs, it would also be interesting to search for localised solutions for these Majorana fermions, in order that they do not combine to form Dirac ones, as one wishes that the neutral excitations could be measured. We will hopefully present such calculations in a future work.

\section*{Acknowledgements}
EMCA would like to thank the kindness and hospitality of Centro Brasileiro de Pesquisas F\' isicas (CBPF), where part of this work was accomplished. EMCA and LPGDA would like to thank CNPq (Conselho Nacional de Desenvolvimento Cient\' ifico e Tecnol\'ogico), a Brazilian Funding Agency, for financial support.

\end{document}